\def\astroph{1}

\ifnum\astroph=0
\documentclass[12pt,preprint]{aastex}       

\else
\documentclass{emulateapj}
\usepackage{apjfonts}
\fi

\usepackage{epsfig,amsmath,natbib}

\newcommand{\fig}[1]{Fig.\ \ref{#1}}
\newcommand{\Fig}[1]{Figure \ref{#1}}
\newcommand{\Eq}[1]{Eq.\ \ref{#1}}

\begin{document}

\shorttitle{Collisionless Shocks}
\shortauthors{C. B. Hededal, T. Haugb\o lle, J. Trier Frederiksen, \AA . Nordlund}

\title{Non--Fermi Power law Acceleration in Astrophysical Plasma Shocks}

\author{C. B. Hededal\altaffilmark{1}, T. Haugb\o lle\altaffilmark{1}, J. Trier Frederiksen\altaffilmark{2},  \AA . 
Nordlund\altaffilmark{1}}
\altaffiltext{1}{Niels Bohr Institute for Astronomy, Physics, and Geophysics, Juliane Maries Vej 30, 2100 
K\o benhavn \O , Denmark}
\altaffiltext{2}{Stockholm Observatory, Roslagstullbacken 21, 106 91 Stockholm, Sweden}
\email{hededal@astro.ku.dk}

\begin{abstract}
Collisionless plasma shock theory, which applies for example to the afterglow of gamma ray
bursts, still contains key issues that are poorly understood. In this paper we
study charged particle dynamics in a highly relativistic collisionless 
shock numerically using $\sim 10^9$ particles.  We find a power law distribution
of accelerated electrons, which upon detailed investigation turns out to originate
from an acceleration mechanism that is decidedly different from Fermi acceleration.
Electrons are accelerated by strong filamentation 
instabilities in the shocked interpenetrating plasmas and coincide spatially with the 
power law distributed current filamentary structures. These structures are an 
inevitable consequence 
of the now well established Weibel--like two--stream instability that operates 
in relativistic collisionless shocks.
The electrons are accelerated and decelerated instantaneously and locally; a scenery 
that differs qualitatively from recursive acceleration mechanisms such as Fermi acceleration.
The slopes of the electron distribution power laws are in concordance with the 
particle power law spectra inferred from observed afterglow synchrotron radiation 
in gamma ray bursts, and the mechanism can possibly explain more generally the 
origin of non--thermal radiation from shocked inter-- and circum--stellar regions
and from relativistic jets.

\end{abstract}


\section{Introduction}
Given the highly relativistic conditions in the outflow from
gamma ray bursts (GRBs), the mean free path for particle Coulomb
collisions in the afterglow shock is several orders of magnitude
larger than the fireball itself.
In explaining the microphysical processes that work to define the
shock, MHD becomes inadequate and collisionless plasma shock theory
stands imperative.
In particular two key issues remain, namely 
the origin and nature of the magnetic field in the shocked region,
and the mechanism by which electrons are accelerated from a thermal
population to a power law distribution $N(\gamma)d\gamma\propto\gamma^{-p}$.
Both ingredients are needed to explain the 
observed afterglow spectra \citep[e.g.][]{2000ApJ...538L.125K, 2001ApJ...560L..49P}.

Regarding the origin of the magnetic field in the shocked region, observations 
are not compatible with a compressed inter--stellar magnetic field, which would
be orders of magnitude smaller than needed \citep{1999ApJ...511..852G}. 
It has been suggested that a Weibel--like two--stream instability can generate a magnetic field in the 
shocked region (\citeauthor{1999ApJ...526..697M} \citeyear{1999ApJ...526..697M}; 
\citeauthor{bib:astro-ph/0303360} \citeyear{bib:astro-ph/0303360}; 
\citeauthor{2003ApJ...595..555N} \citeyear{2003ApJ...595..555N}; 
\citeauthor{2003ApJ...596L.121S} \citeyear{2003ApJ...596L.121S}).
Computer experiments by \citet{2004ApJ...608L..13F} showed that the nonlinear stage of a 
two--stream instability induces a magnetic field {\it in situ} with an energy content of a
few percent of the equipartition value, consistent with
that required by observations. 

Fermi acceleration \citep{1949PhRv...75.1169F} has, so far,
been widely accepted as the mechanism that provides the inferred electron 
acceleration. 
It has been employed extensively in Monte Carlo simulations 
(e.g.~\citet{bib:Niemiec} and references therein),
where it operates in conjunction with certain assumptions about the scattering of particles
and the structure of the magnetic field.
The mechanism has, however,
not been conclusively demonstrated to occur in {\em ab initio} particle simulations.
As pointed out by \citet{bib:Niemiec}, 
further significant advance in the study of relativistic shock particle acceleration is
unlikely without understanding the detailed microphysics of collisionless shocks. Also,
recently \citet{bib:baring} found that particle distribution functions (PDFs)
inferred from GRB observations are in contradistinction with standard
acceleration mechanisms such as diffusive Fermi acceleration. 

In this letter we study {\em ab initio} the particle dynamics 
in a collisionless shock with bulk Lorentz factor $\Gamma=15$.
We find a new particle 
acceleration mechanism, which we present in Section 2.  Detailed 
numerical results are presented 
and interpreted in Section 3, while Section 4 contains the conclusions.
\begin{figure*}[!th]
\begin{center}
\epsfig{figure=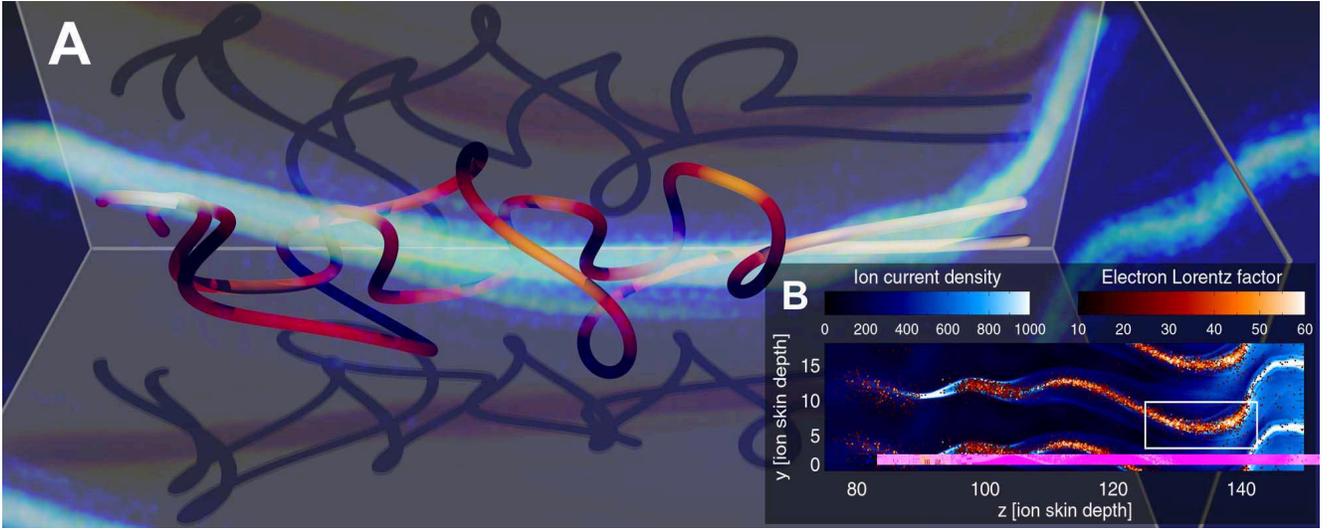,width=17.5cm}
\caption{(A) Ray traced electron paths (red) and current density (blue).
The colors
of the electron paths reflect their four velocity according to the color table in the inset (B). The shadows are
equivalent to the $x$ and $y$ projections of their paths. The ion current density is
shown with blue colors according to the color table in the inset. The inset also
shows the ion current density (blue) integrated along the
$x$ axis with the spatial distribution of fast moving electrons (red) over plotted.}
\label{fig:acceleration}
\end{center}
\end{figure*}

\section{A new acceleration mechanism}
We have performed a series of numerical experiments where collisionless 
shocks are created by two colliding plasma populations.  These experiments 
are described in more detail below, but a common feature is 
that the electron PDF has a high energy tail which is power law distributed.  By 
carefully examining the paths of representative accelerated electrons, 
tracing them backwards and forwards in time, we have been able to 
identify the mechanism responsible for their acceleration.
The acceleration mechanism, which as far as we can tell has not been 
discussed in the literature previously, works as follows:

When two non--magnetized collisionless plasma populations interpenetrate, current channels are formed 
through a Weibel--like two--stream instability 
(\citeauthor{1999ApJ...526..697M} \citeyear{1999ApJ...526..697M};
\citeauthor{bib:astro-ph/0303360} \citeyear{bib:astro-ph/0303360};
\citeauthor{2003ApJ...595..555N} \citeyear{2003ApJ...595..555N};
\citeauthor{2003ApJ...596L.121S} \citeyear{2003ApJ...596L.121S}).
In the nonlinear stage of evolution of this instability, ion current channels merge 
into increasingly stronger patterns, while electrons act to 
Debye shield these channels, as shown by \citet{2004ApJ...608L..13F}. 
That work further showed
that a Fourier decomposition of the transverse ion current filaments 
exhibits power law behavior which has been recently confirmed by \citet{bib:medv}.

At distances less than the Debye length, the ion current channels are surrounded 
by transverse electric fields that accelerate the electrons toward the current 
channels. However, the magnetic fields that are induced around the current channels 
act to deflect the path
of the accelerated electrons, boosting them instead in the direction of the ion flow.
Since the forces working are due to quasi--stationary fields the acceleration is a
simple consequence of potential energy being converted into kinetic energy. Therefore
the electrons are decelerated again when leaving the current channel, and reach
their maximal velocities at the centers of the current channels. Hence, as illustrated by
\fig{fig:acceleration}B, the spatial distribution
of the high energy electrons is a direct match to the ion current channels and 
the properties of the accelerated electrons depend
primarily on the local conditions in the plasma.

One might argue that the near--potential behavior of the electrons, where they essentially
must loose most of their energy to escape from the current channels, would make the
mechanism uninteresting as an acceleration mechanism since fast electrons cannot easily
escape.  However, this feature may instead be a major advantage, since it means that
energy losses due to escape are small, and that the electrons remain trapped long enough
to have time to loose their energy via a combination of bremsstrahlung and synchrotron  
or jitter radiation.
We observe that only a very small fraction of the electrons manage to escape,
while still retaining most of their kinetic energy. 
This happens mainly at sudden bends or mergers of
the ion channels, where the electron orbits cannot be described in terms of 
a particle moving in a static electromagnetic field.

\begin{figure}[!ht]
\begin{center}
\epsfig{figure=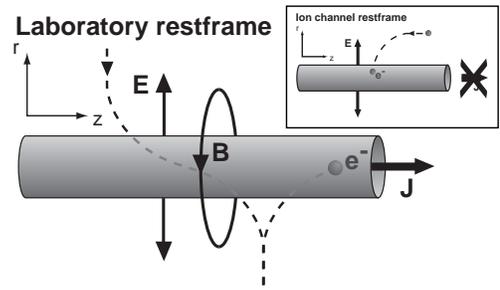,width=6.5cm}
\caption{An ion current channel surrounded by an electric -- and a magnetic field.
Electrons in the vicinity of the current channels are thus subject to a Lorentz force
with both an electric and magnetic component, working together to accelerate the
electrons along the ion flow. Crossing the center of the channel the process
reverses leading to an oscillating movement along the channel.}
\label{fig:current_acc}
\end{center}
\end{figure}

To analyze the acceleration scenario quantitatively 
we use the sketch in \fig{fig:current_acc}.
We assume that the ion current channel has radius $R$, that the total charge inside
the cylinder per unit length is $\lambda$ and the ions stream with velocity
$u$ in the laboratory rest frame
(see \fig{fig:current_acc} and inset for definition of rest frames).
Consider an electron
with charge $-q$ and mass $m$ at a distance $r$ from the center of the channel,
initially having no velocity components perpendicular to the cylinder.
By analyzing everything in the ion channel rest frame the problem reduces
to electrostatics and it is possible to analytically calculate the change in 
four velocity of the electron when it reaches the surface of the cylinder.
Since the electric force only works along the $r$-axis, the four velocity
along the $z$--axis of the electron is conserved in the ion channel rest frame.
Hence we can calculate both the total change in energy and the change in
the different velocity components.
Returning to the laboratory rest frame we find
\begin{align}\label{eq:acc}
\Delta\gamma_{electron} &= \frac{q \lambda}{2 \pi m c^2 \epsilon_0} 
             \ln \frac{r}{R} \\
\Delta(\gamma v_z )_{electron} &= u \Delta\gamma_{electron}\, .
\end{align}
The change in the Lorentz boost is directly proportional to
the total charge inside the channel
and inversely
proportional to the electron mass. Debye shielding 
reduces the electric field further away from the ion channel, so the estimate
above is only valid for distances smaller than a Debye length. 

\section{Computer Experiments}  
The experiments were performed with the
three dimensional relativistic kinetic and electromagnetic particle--in--cell code 
described by \citet{2004ApJ...608L..13F}. The code works from first principles, 
by solving Maxwell's equations for 
the electromagnetic fields and solving the Lorentz force equation of motion for the particles.

Two colliding plasma populations are set up in the rest frame of one 
of the populations (downstream, e.g. a jet). A less dense population (upstream,
e.g. the ISM) is continuously injected at the left boundary with a relativistic velocity corresponding
to a Lorentz factor $\Gamma=15$. The two populations initially differ in density by a factor of 3.
We use a computational box with $125\times125\times2000$ grid points and a
total of $8\times10^8$ particles. The ion rest frame plasma frequency in the downstream
medium is $\omega_{pi}=0.075$, rendering the box 150 ion skin depths long.
The electron rest frame plasma frequency is $\omega_{pe}=0.3$ in order to resolve
also the microphysics of the electrons.
Hence the ion-to-electron mass ratio is $m_i/m_e = 16$. Other mass ratios and
plasma frequencies were used in complementary experiments. 
Initially, both plasma populations are unmagnetized.

The maximum experiment duration has $t_{max} =$ 340 $\omega_{pi}^{-1}$, which 
is sufficient for the continuously injected upstream plasma ($\Gamma = 15$, $v\sim c$)
to travel 2.3 times the length of the box.
The extended size and duration of these experiments enable
observations of the streaming instabilities and concurrent particle 
acceleration through several stages of development \citep{2004ApJ...608L..13F}. 
Momentum losses to radiation (cooling) are presently not included in the 
model. We have, however, verified that none of the accelerated particles in the
experiment would be subject to significant synchrotron
cooling.   The emitted radiation may thus be
expected to accurately reflect the distribution of accelerated electrons.

When comparing numerical data with \Eq{eq:acc}
we take $r$ to be the radius where Debye shielding starts to be
important. Using a cross section
approximately in the middle of \fig{fig:acceleration}
we find $\Delta(\gamma v_z)_{electron} = 58 \ln (r/R)$. It is hard to determine
exactly when Debye shielding becomes effective, but looking at electron paths 
and the profile of the electric field we estimate that
$\ln (r/R) \approx 1.3$. Consequently, according to \Eq{eq:acc}, the 
maximally attainable four velocity in this experiment is in the neighborhood of 
$(\gamma v_z)_{max}=75$. This is in good agreement with the results from our 
experiments, where the maximum four velocity is $(\gamma v_z)_{max}\simeq80$.

\begin{figure}[!t]
\begin{center}
\epsfig{figure=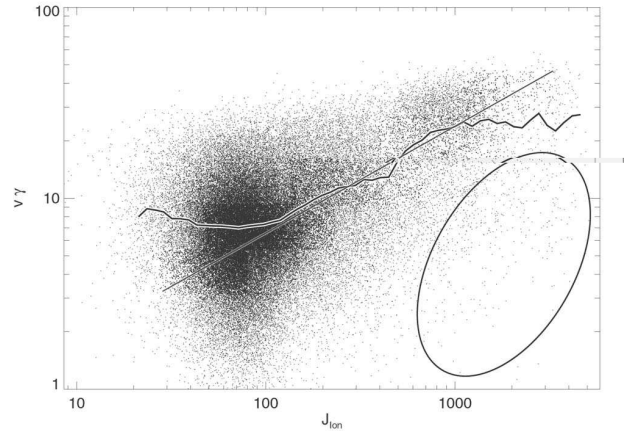,width=8.5cm}
\caption{A scatter plot of the local ion current density $J_{Ion}$ versus the four velocity
of the electrons in a region downstream of the shock. Overplotted is a line (thin) showing the
average four velocity as a function of $J_{Ion}$, and a line (thick) showing a straight line fit.
Because 'cold' trapped thermal electrons (indicated
with the ellipse) exist inside the ion current channel they count towards lowering the average
four velocity at high $J_{Ion}$. If we cleaned our scatter plot, statistically removing all
thermal electrons we would see a much tighter relation. Such cleaning, though, is rather
delicate and could introduce biases by itself. The trend is clearly there though even for the
'raw' data.}
\label{fig:jvg}
\end{center}
\end{figure}

The theoretical model does of course not cover all details of the experiment. 
For example, in general the
electrons also have velocity components parallel to the magnetic field; instead of
making one dimensional harmonic oscillations in the plane perpendicular to the current
channel the electrons will describe ellipsoidal paths.
\fig{fig:acceleration} shows the path of two electrons in the vicinity of an ion channel. But, overall, the electrons behave as expected from the model considerations. Consequently, high speed electrons are tightly coupled to the ion channels,
as clearly illustrated by \fig{fig:acceleration}B.

\begin{figure}[!t]
\begin{center}
\epsfig{figure=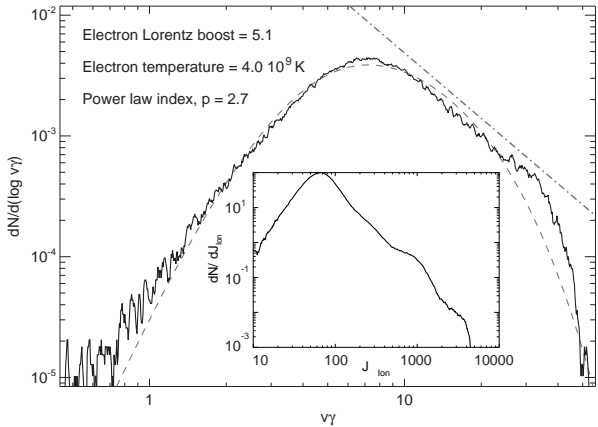,width=8.5cm}
\caption{The normalized electron particle distribution function downstream of
the shock. The dot--dashed line is a power law fit to the non--thermal high
energy tail, while the dashed curve is a Lorentz boosted thermal electron
population. The histogram is made from the four velocities of electrons in a thin slice in the $z$--direction
of the computational box. The inset shows a similar histogram for ion current density
sampled in each grid point in the same slice. The bump in the inset is a statistical fluctuation due to a single
ion channel.
}
\label{fig:pdfpower}
\end{center}
\end{figure}

\Fig{fig:pdfpower} shows that the electrons are power law distributed at 
high energies, with index $p=2.7$. 
The electrons at the high gamma cutoff are found where the ion current peaks, 
as may be seen from \fig{fig:jvg}. The maximum ion current is limited 
by the size of our box; larger values would probably be found if the 
merging of current channels could be followed further down stream.
The PDF is not isotropic in any frame of reference due to the high 
anisotropy of the Weibel generated electromagnetic field.
The power law in the electron PDF is dominant for $10<\gamma<30$. 
Likewise, a power law dominates the ion current channel strength, $J_{Ion}$, for 
$100<J_{Ion}<1000$ (inset). 
A relation between the power law distributions of these two quantities on their respective 
intervals is provided with \fig{fig:jvg}: We see that the average four velocity is proportional 
(straight line fit) to a power of the local ion current density on their respective relevant intervals, 
$10<\gamma<30$ and $100<J_{Ion}<1000$. Their kinship stems from the fact that acceleration is
local. $J_{Ion}$ has a power law tail and its potential drives the high energy distribution of
the electrons according to \Eq{eq:acc}, thus forming a power law distributed electron PDF.

Measuring the rate at which the in--streaming ions transfer momentum to the 
ion population initially at rest allows us to make a crude estimate of the length scales
over which the two--stream instability in the current experiment would saturate due to ion
thermalization. A reasonable estimate appears to be approximately 10 times
the length of the current computational box, or about 1500 ion skin depths. Assuming that
the shock propagates in an interstellar environment with a plasma density of 
$\sim 10^6$ m$^{-3}$ we may calculate a typical
 ion skin depth. Comparing this value with the upstream ion skin depth from our experiments, 
we find that the computational box corresponds to a scale of the order  of $10^7$ m,
or equivalently that the collisionless shock transition 
region of the current experiment corresponds to about $10^8$ m.
For an ion with a Lorentz factor $\gamma=15$ this length corresponds roughly
to 40 ion gyro radii in the average strength of the generated magnetic field. 
But we stress that the in--streaming ions actually do not really gyrate 
since they mainly travel inside the ion current channels where the magnetic 
field, by symmetry, is close to zero. Also, the strong electromagnetic fields 
generated by the Weibel instability and the non-thermal electron acceleration, 
which is crucial from the interpretation of GRB afterglow observations,
emphasize the shortcoming of MHD in the context of collisionless shocks.

In the computer experiments presented here we have used an mass ratio 
$m_i/m_e=16$ in order to resolve the dynamics of both species. \Eq{eq:acc} suggests that
reducing the electron mass to $1/1836\,m_i$ will increase the acceleration of 
the electrons to maximum energies in the neighborhood of 5 GeV. 
Even further acceleration may occur as ion channels keep growing down stream,
outside of our computational box.

The scaling estimates above depend, among other things, on plasma 
densities, the bulk Lorentz factor, and the mass ratio ($m_i/m_e$).
A parameter study is necessary to explore these dependencies,
but this is beyond the scope of the present paper. We thus stress that the
 extrapolations performed here are speculative and that 
unresolved physics could influence the late stages
of the instability in new and interesting ways. 

When the in--streaming ions are fully thermalized they can no longer support the 
magnetic field structures. Thus one might speculate that the radiating region of 
the GRB afterglow is actually very thin, as  suggested by \citet{2003MNRAS.339..881R}. 
Further, traditional synchrotron radiation theory does not apply to intermittent 
magnetic field generated by the two--stream instability, since the 
electron gyro radii often are larger than the scales of the magnetic field structures. 
We emphasize the importance of the theory of 
jitter radiation \citep{2000ApJ...540..704M}.

\section{Conclusions}
We have proposed a new acceleration mechanism for electrons in
collisionless shocks. The theoretical considerations were suggested by
particle--in--cell computer experiments, which also allowed quantitative 
comparisons with the theoretical predictions. We have shown that the 
non--thermal acceleration of electrons is directly related to the 
ion current channels in the shock transition zone.
The results are applicable to
interactions between relativistic outflows and the interstellar medium.
Such relativistic outflows occur in GRB afterglows and in jets from
compact objects \citep{2004Natur.427..222F}. The suggested acceleration
scenario might overcome some of the problems pointed out by \citet{bib:baring}
regarding the apparent contradiction between standard Fermi acceleration and 
spectral observations of GRBs.

The mechanism has important implications for the way 
we understand and interpret observations of collisionless shocks:

1. The acceleration mechanism is capable of creating a power law 
electron distribution in a collisionless shocked region. 
In the computer experiment presented here a bulk flow with
$\Gamma=15$ results in a power law slope $p=2.7$ for the electron PDF.
Additional experiments will be needed to disentangle what determines
the exact value of the slope.

2. The acceleration is local; electrons
are accelerated to a power law in situ. 
Therefore the observed radiation field may 
be tied directly to the local conditions of the plasma and could be a strong 
handle on the physical processes.

3. Our results strengthen the point already made by \citet{2004ApJ...608L..13F};
that the fractions of the bulk kinetic energy that go into in the electrons 
and the magnetic field, $\epsilon_e$ and $\epsilon_B$ respectively, are not 
free and independent parameters
of collisionless shock theory. Most likely they represent interconnected parts
of the same process. 

4. In the case of a weak or no upstream magnetic field, the Weibel--like two--stream
instability is able to provide the necessary electromagnetic fields. We have shown here
that the collisionless shocked region is relatively thin, and we suggest that the
non--thermal radiation observed from GRB afterglows 
and relativistic jets in general is emitted from such a relatively thin shell. 

It is clear that the non-thermal electron acceleration, the ion current filamentation, the magnetic field amplification/generation, and hence the strong non-thermal radiation from the shock, is beyond the reach of MHD to explain. Whether the relativistic MHD jump conditions become valid on any larger scale is not possible to decide from the simulations presented in this paper.

We thank the Danish Center for Scientific Computing for granting the 
computer resources that made this work possible. The authors will also
like to thank Ken-Ichi Nishikawa, Mikhail Medvedev and the referee for their 
comments and suggestions.  The work of {\AA}N
was supported by a grant from the Danish Natural Science Research Council.

\bibliographystyle{apj}
\bibliography{particles}

\end{document}